\documentclass[runningheads]{llncs}

\usepackage{graphicx} %
\usepackage{mathtools}
\usepackage{amssymb} %
\usepackage{subfig}
\usepackage{xspace}
\usepackage{amsfonts}
\usepackage[ruled,vlined,linesnumbered]{algorithm2e}
\usepackage[bibliography=common]{apxproof}

\usepackage{tikz}
\usepackage{pgfplots}
\pgfplotsset{compat=newest}
\usetikzlibrary {automata,positioning,fit,calc,patterns}
\usepgfplotslibrary{statistics}

\DeclarePairedDelimiter{\parentheses}{(}{)}

\NewDocumentCommand{\Lang}{}{\mathcal{L}\parentheses}
\NewDocumentCommand{\Fin}{}{\operatorname{Fin}\parentheses}
\NewDocumentCommand{\Inf}{}{\operatorname{Inf}\parentheses}

\newcommand{\Aut}{\mathcal{A}}
\newcommand{\States}{Q}
\newcommand{\AP}{\Sigma}
\newcommand{\Acc}{\textit{acc}}
\DeclarePairedDelimiter\Angle{\langle}{\rangle}

\newtheoremrep{apxlemma}{Lemma}
\newtheoremrep{apxtheorem}{Theorem}

\newcommand*{\HOAexec}{\textsc{Hoax}\xspace}
\newcommand*{\Sccs}{\mathbb{S}}

\begin{document}

\title{Execution and monitoring of HOA automata with HOAX}
\subtitle{(Extended version)}
\author{Luca Di Stefano\orcidID{0000-0003-1922-3151}}
\authorrunning{L. Di Stefano}

\institute{%
TU Wien, Institute of Computer Engineering, Treitlstraße 3, 1040 Vienna, Austria\\
\email{luca.di.stefano@tuwien.ac.at}%
}

\maketitle

\begin{abstract}
We present a tool called \HOAexec for the execution of $\omega$-automata expressed in
the popular HOA format. The tool leverages the notion of trap sets to enable runtime monitoring of any (non-parity) acceptance condition supported by the format. When the automaton is not monitorable, the tool may still be able to recognise so-called ugly prefixes, and determine that no further observation will ever lead to a conclusive verdict.
The tool is open-source and highly configurable. We present its formal foundations, its design, and compare it against the trace analyser PyContract on a lock acquisition scenario.
\end{abstract}

\section{Introduction}

The Hanoi Omega-Automata (HOA) format is a well-established language to describe and exchange $\omega$-automata~\cite{DBLP:conf/cav/BabiakBDKKM0S15}, featuring a rich grammar for acceptance conditions that generalises (co-)B\"uchi, Strett, Rabin, and other classical  families of automata.
HOA is supported by mature toolboxes, including
Spot~\cite{DBLP:conf/atva/Duret-LutzLFMRX16}, PRISM~\cite{DBLP:conf/cav/KwiatkowskaNP11}, and Owl~\cite{DBLP:conf/atva/KretinskyMS18},
which provide valuable tools to construct and analyse automata.
However, there are no tools to \emph{execute} an automaton, i.e., to display how it evolves through its state space under a sequence of inputs.

In this work, we present an open-source utility to fill this gap, called \HOAexec.\footnote{Short for \textsc{Hoa} e\textsc{x}ecutor.}
It receives as input one or more HOA files, and executes the automata described therein, by reading input valuations from a configurable set of sources.
It also allows customized behaviour when, for instance, nondeterminism or deadlock are detected in the automaton, or when user-defined conditions are met.

Runtime verification represents a natural application of this tool.
To enable this, we implement within \HOAexec a reasoning mechanism based on the notion of \emph{trap sets}~\cite{DBLP:journals/nc/KlarnerBS15}, i.e., subsets of the state space that the automaton cannot leave once entered.
This reasoning lets the tool monitor deterministic complete automata with any of the (non-parity) arbitrary acceptance conditions expressible in HOA. If a finite sequence of inputs is enough to deduce that the condition will be fulfilled or not, the tool reports a conclusive verdict. Although this is not guaranteed to happen in finite time,
if the automaton is not monitorable \HOAexec can also try to detect ugly prefixes~\cite{DBLP:journals/tosem/BauerLS11}, to decide whether further monitoring is useful.

This paper is structured as follows. Section~\ref{sec:background} contains definitions that will be used throughout the paper.
Section~\ref{sec:trap} discusses how trap sets may be computed and used for monitoring.\footnote{Proofs for the lemmas and theorems in this section are reported in the appendix.}
In Section~\ref{sec:tool} we describe the workflow and features of \HOAexec, and compare it with
PyContract~\cite{DBLP:conf/rv/DamsHK22}, a Python library for building trace analysers.
Lastly, we make our concluding remarks in Section~\ref{sec:conclusion}.

\noindent\emph{Related work.}
Several formats for graphs and automata support, unlike HOA, some form of execution or simulation.
For instance,
the CADP suite~\cite{DBLP:journals/sttt/GaravelLMS13} contains tools for the (random or interactive) execution of transition systems expressed either explicitly or through a process calculus such as LNT~\cite{DBLP:conf/birthday/GaravelLS17}.
mCLR2~\cite{DBLP:conf/tacas/BunteGKLNVWWW19} and the TLC toolbox for TLA+~\cite{DBLP:conf/charme/YuML99} provide similar utilities for their respective formalisms.
Since these formalism target transition systems, rather than automata,
they do not usually incorporate acceptance conditions.
Similarly, the \texttt{aigsim} tool can simulate a circuit expressed in the AIGER format~\cite{biere2007}. Inputs may be either random or read from a stimulus file, whereas \HOAexec can mix different sources of input.
In turn, AIGER partitions the alphabet of the circuit into \emph{inputs} and \emph{outputs}, making the format amenable to describe SAT and model checking queries, as well as reactive synthesis problems.
PyContract~\cite{DBLP:conf/rv/DamsHK22} is a Python library by which monitors can be explicitly programmed in a shallow domain-specific language (DSL).
These monitors are symbolic, as they observe data-carrying messages
rather than plain Boolean signals; this, in turn, enables advanced techniques, such as slicing, to further improve performance. While the library is oriented towards trace analysis, its API may be leveraged to also implement online monitoring.
Spot lets one build a HOA automaton that monitors a property $\varphi$
from the safety fragment of Linear Temporal Logic (LTL~\cite{DBLP:conf/focs/Pnueli77}). The resulting automaton may deadlock on specific inputs, and finding a deadlock corresponds to detecting a violation of~$\varphi$.\footnote{See \url{https://spot.lre.epita.fr/tut11.html}.}
Our tool may be instrumented to support these automata, and its reasoning extends to full LTL, although with no guarantees that a conclusive verdict will be reached.

\section{Background}\label{sec:background}

\begin{definition}[Words, prefixes, suffixes]
Given a finite set $\AP$ (also called an \emph{alphabet}),
we denote by $\AP^\omega$ the set of (infinite) \emph{words} over $\AP$.

For any word $w=w_0 w_1 w_2\ldots$ and any $i \geq 0$, we can always split $w$ into a \emph{prefix}
$w_{\leq i} = w_0 w_1 w_2 \dots w_i$
of finite length $i+1$,
and an infinite \emph{suffix} $w_{> i} = w_{i+1} w_{i+2} \dots$; we denote by $\AP^\ast$ the set of all word prefixes, or \emph{finite words}.
\end{definition}

\begin{definition}[Automata]
An \emph{automaton} is a tuple
\(\Aut = \Angle*{\States, \AP, \mathord{\rightarrow}, \States_0, \varphi}
\)
where $\States$ is a finite set of \emph{states},
$\AP$ a finite \emph{alphabet},
$\mathord{\rightarrow}\subseteq \States\times\AP\times\States$ a labelled \emph{transition relation},
$\States_0 \subseteq \States$ a non-empty set of \emph{initial states},
and $\varphi$ an \emph{acceptance condition} from \( \Acc_\States\), defined as:

\[
\Acc_\States \coloneqq
\top 
\mid \bot 
\mid \Fin{S}
\mid \Inf{S}
\mid \Acc_\States \land \Acc_\States
\mid \Acc_\States \lor \Acc_\States,
\]
where $S$ denotes non-empty subsets of $\States$.
We write $q \xrightarrow{a} q'$ iff $(q, a, q') \in \mathord{\rightarrow}$, and 
$q \rightarrow q'$ iff $(q, a, q') \in \mathord{\rightarrow}$ for some $a$.
\end{definition}

\begin{definition}[Deterministic and complete automata]
    An automaton $\Aut = \Angle*{\States, \AP, \mathord{\rightarrow}, \States_0, \varphi}$ is \emph{deterministic} iff i) $|\States_0| = 1$, and ii) for every pair $(q, a) \in \States \times \AP$ there is at most one $q'\in \States$ such that $q\xrightarrow{a}q'$.
    An automaton is \emph{complete} if for every pair $(q, a) \in \States \times \AP$ there is at least one $q'$ such that $q\xrightarrow{a}q'$.
\end{definition}

\newcommand{\AAPP}{\textit{AP}}
Note that the alphabet of a HOA automaton~\cite{DBLP:conf/cav/BabiakBDKKM0S15} is the set of Boolean formulas over a finite set of \emph{atomic propositions} $\AAPP$, denoted by $\mathbb{B}(\AAPP)$.\footnote{%
This set is infinite, but its quotient modulo equivalence of truth tables is finite with size $2^{2^{|\AAPP|}}$, and thus we may treat it as an alphabet.}
Two formulas in $\mathbb{B}(\AAPP)$ are \emph{disjoint} iff there is no valuation of $\AAPP$ that satisfies both.
A HOA automaton is deterministic iff it has one initial state and, in every state, outgoing transitions are labelled by pairwise disjoint formulas; and it is complete iff, in every state, the disjunction of all labels on outgoing transitions is a valid formula.

\begin{definition}[Runs]
A sequence of states $\rho = \rho_0 \rho_1 \rho_2 \dots$
is a \emph{run} in $\Aut$ for a word $w=w_0 w_1 w_2\dots$
if $\rho_0 \in \States_0$ and, for every $i\geq 0$,
$\rho_i \xrightarrow{w_i} \rho_{i+1}$.
Similarly to what we did with words, we denote
by $\rho_{\leq i}$ and $\rho_{> i}$ the prefix and suffix of a run.
We write $w_\Aut(\rho)$ to denote the word for which $\rho$ is a run in $\Aut$.

\end{definition}

\begin{definition}[Accepting runs, languages]\label{def:accept}
A run \(\rho\) \emph{models} a condition $\varphi$,
denoted by $\rho \models \varphi$,
according to the following interpretation:
\begin{align*}
   \rho &\models \top
   &
   \rho &\models \Fin{S}\Leftrightarrow \exists i. \forall j. j>i \Rightarrow \rho_j\not\in S
   &
   \rho &\models \varphi_1 \land \varphi_2 \Leftrightarrow
   \rho \models \varphi_1 \text{ and }
   \rho \models \varphi_2
   \\   
   \rho &\not\models \bot
   &
   \rho &\models \Inf{S}\Leftrightarrow \rho\not\models\Fin{S}
   &
   \rho &\models \varphi_1 \lor \varphi_2 \Leftrightarrow
   \rho \models \varphi_1 \text{ or }
   \rho \models \varphi_2
\end{align*}

A run that models $\varphi$ is also called an \emph{accepting run} for it.
The \emph{language} \(\Lang{\Aut}\) of an automaton \(\Aut\) with acceptance condition $\varphi$ is the set of words $w \in \AP^\omega$ for which there exists an accepting run for $\varphi$ in \(\Aut\).
\end{definition}

Intuitively, a run $\rho$ is accepting for $\Fin{S}$ (and not accepting for $\Inf{S}$) iff
it only visits $S$ a finite number of times.
For instance, a run of an automaton with a transition relation like the one in Fig.~\ref{fig:a1} starts
in $q_0$, immediately leaves it, and can never visit $q_0$ again. Therefore, the run models
$\Fin{\{q_0\}}$.

Furthermore, we will call a \emph{transient} any set of states $S$ of an automaton $\Aut$ such that
every run of $\Aut$ models \( \Inf{Q\setminus S}\).
Notice that a transient may still be visited infinitely often during a run:
the only requirement is that every run also \emph{leaves} it infinitely often.
To determine whether $S$ is a transient, it suffices to check that the transitions $\{q \rightarrow q' \mid q,q' \in S\}$ are acyclic, which can be done in
$\mathcal{O}(|S| + |\mathord{\rightarrow}|)$ time in the worst case.

\begin{definition}[Strongly connected components; Bottom SCCs]\label{def:bscc}
    A strongly connected component (SCC) of a directed graph
    is any maximal set of pairwise-reachable vertices in the graph.
    A bottom SCC (BSCC) is an SCC that is closed under the edge relation of the graph.
\end{definition}

We will extend these graph-theoretical definitions to an automaton with states $\States$ and
transition relation $\rightarrow$ by considering the directed graph that has vertices $\States$
and has an edge $(q, q')$ if and only if $q \rightarrow q'$.

\begin{definition}[Good, bad, ugly prefixes; monitorability~\cite{DBLP:journals/fmsd/KupfermanV01,DBLP:journals/tosem/BauerLS11}.]
    Let \(\Aut\) an automaton over alphabet $\AP$, and $u \in \AP^\ast$ a finite word. We say that $u$ is
    \begin{itemize}
        \item a \emph{good} prefix iff, for all $w \in \AP^\omega$, $uw \in \Lang{\Aut}$;
        \item a \emph{bad} prefix iff, for all $w \in \AP^\omega$, $uw \not\in \Lang{\Aut}$;
        \item an \emph{ugly} prefix iff, for all $v \in \AP^\ast$, $uv$ is neither good nor bad.
    \end{itemize}
We say that $\rho_{\leq i}$ is a good, bad, or ugly run prefix if
$(w_\Aut(\rho))_{\leq i}$ is a good, bad, or ugly prefix, respectively.
An automaton is \emph{monitorable} if it has no ugly prefixes.
\end{definition}

\begin{definition}[Trap sets~\cite{DBLP:journals/nc/KlarnerBS15}]\label{def:trap}
    Let \(\Aut = \Angle*{\States, \AP, \mathord{\rightarrow}, \States_0, \varphi}\)
    an automaton.
    Then, a non-empty set \(T\subseteq \States\) is a \emph{trap set} of \(\Aut\) if, for every \(q\in T\), $q\rightarrow q'$ implies \(q' \in T\).
    A trap set $T$ is \emph{minimal} iff there exists no subset $T' \subset T$ that is also a trap set.
    $\States$ is always a \emph{trivial} trap set for \(\Aut\).
\end{definition}

\begin{lemma}\label{lem:trap}
\(T\) is a trap set of a complete automaton \(\Aut\) iff
for every run $\rho$ of \(\Aut\),
if $\rho_i \in T$ for some $i\geq 0$, then $\rho_j \in T$ $\forall j \geq i$.
\end{lemma}
\begin{inlineproof}
($\Rightarrow$) By induction. If $j=i$, we already know that $\rho_i \in T$. If $j>i$, assume that $\rho_k \in T$ for $k = i, i+1, \ldots, j-1$. 
By definition of a run, $\rho_{j-1} \rightarrow \rho_{j}$.
Then, by definition of $T$, \(\rho_{j}\) also belongs to $T$.
($\Leftarrow$)
If every run $\rho$ entering $T$
stays within $T$ forever, this means that
there cannot exist any $q \not\in T$ such that $q \rightarrow q'$. This means that $T$ is closed under $\rightarrow$ and thus a trap set.
\qed
\end{inlineproof}

\section{Runtime monitoring with trap sets}\label{sec:trap}

In this section, we describe how to compute the trap sets of an automaton (technically, of its underlying state-transition graph); then, we exploit
them to monitor an acceptance condition as defined in (Def.~\ref{def:accept})
on a complete deterministic automaton.
When appropriate, we can use the same technique on an incomplete automaton that we make complete by adding adequate stuttering transitions.%

\noindent\emph{Computing trap sets.}
The trap sets of an automaton have a nuanced relation with its strongly connected components.
For instance, Fig.~\ref{fig:a1} shows a graph with trap sets \(\{q_1, q_2\}\) and \(\{q_2\} \) (which is minimal), where the former is not an SCC.
At the same time, in Fig.~\ref{fig:a2}, sets \(\{s_0, s_1\}\) is an SCC but not a trap set, since it is not closed under $\rightarrow$. By contrast, $\{s_2\}$ is both an SCC and a trap set.
We now characterise how trap sets and SCCs relate to one another:

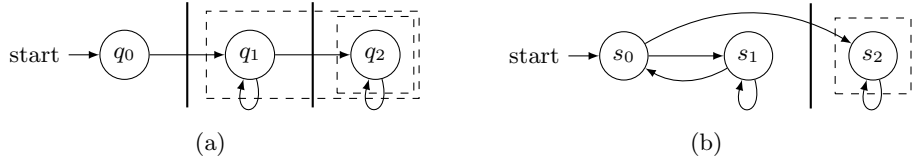
\begin{figure}[tb]
\centering
\subfloat[\label{fig:a1}]{%
\begin{tikzpicture}[auto, >=latex]

\begin{scope}[font=\footnotesize]
\node[state,initial,minimum size=0pt]  (q_0) {$q_0$};
\node[state,minimum size=0pt]          (q_1) [right=of q_0] {$q_1$};
\node[state,minimum size=0pt]          (q_2) [right=of q_1] {$q_2$};
\node (x1) at ($(q_0)!0.5!(q_1)$) {};
\node (x2) at ($(q_1)!0.5!(q_2)$) {};
\end{scope}

\node[draw,dashed,fit=(q_1) (q_2),inner sep=7pt] (ts1) {};
\node[draw,dashed,fit=(q_2) (q_2),inner sep=5pt] (ts2) {};

\draw[->] (q_0) -- (q_1) ;
\draw[->] (q_1) -- (q_2) ;
\draw[->] (q_1) to[loop below] (q_1);
\draw[->] (q_2) to[loop below] (q_2);

\draw[thick] ($(x1)-(0,.7)$) -- ($(x1)+(0,.7)$);
\draw[thick] ($(x2)-(0,.7)$) -- ($(x2)+(0,.7)$);
\end{tikzpicture}}\hspace{3em}
\subfloat[\label{fig:a2}]{%
\begin{tikzpicture}[auto, >=latex]
\begin{scope}[font=\footnotesize]
\node[state,initial,minimum size=0pt]  (q_0) {$s_0$};
\node[state,minimum size=0pt]          (q_1) [right=of q_0] {$s_1$};
\node[state,minimum size=0pt]          (q_2) [right=of q_1] {$s_2$};
\node (x1) at ($(q_0)!0.5!(q_1)$) {};
\node (x2) at ($(q_1)!0.5!(q_2)$) {};
\end{scope}

\node[draw,dashed,fit=(q_2) (q_2),inner sep=5pt] (ts2) {};

\draw[->] (q_0) -- (q_1) ;
\draw[->] (q_1) to[bend left] (q_0) ;

\draw[->] (q_0) to[bend left] (q_2) ;
\draw[->] (q_1) to[loop below] (q_1);
\draw[->] (q_2) to[loop below] (q_2);

\draw[thick] ($(x2)-(0,0.7)$) -- ($(x2)+(0,0.7)$);

\end{tikzpicture}}
\caption{Examples of graphs with their trap sets (enclosed in dashed boxes) and SCCs (separated by solid lines).}\label{fig:scc-example}
\end{figure}

\begin{apxlemmarep}\label{lem:bscc1}
    Let $\Aut$ a complete automaton and let $T$ a subset of its states.
    $T$ is a BSCC of $\Aut$ iff it is a minimal trap set of \(\Aut\).
\end{apxlemmarep}
\begin{proof}
    ($\Rightarrow$) By Defn.~\ref{def:bscc}, a BSCC is closed under the transition relation, i.e., has no outgoing transitions.
    This makes it also a trap set.
    ($\Leftarrow$) Let $T$ a minimal trap set. By definition of trap set, it is closed under reachability. We have to prove that $T$ is also an SCC, that is, every state is reachable from any other state. 
    Therefore, let us assume that $T$ is \emph{not} an SCC:
    that is, there exists a state $q \in T$ from which one may not
    reach $T \setminus \{q\}$.
    This state must have at least one outgoing transition $q\rightarrow q'$, otherwise the singleton set $\{q\}$ would be
    a trap set and $T$ would not be minimal.
    But if $q'\neq q$ then $q' \in T$, by definition of trap set.
    Conversely, if $q' = q$, and there are no other transitions from $q$, then we may indeed never visit $T\setminus \{q\}$ again. But this means that $T$ is not minimal: in fact, it contains a smaller trap set which is \(\{q\}\) itself.
    Both cases are contradictory.\qed
\end{proof}

\begin{apxlemmarep}\label{lem:scc}
Let $K$ an SCC of $\Aut$ that is not a bottom SCC.
Denote by $N(K)$ the set of states $n \not\in K$ such that $q\rightarrow n$ for some $q \in K$.
If every such state belongs to a trap set $T_n$, then the set $T' = K \cup \bigcup_{n\in N(K)} T_n$ is also a trap set.
\end{apxlemmarep}
\begin{proof}
    Let $\rho$ any run of $\Aut$ such that $\rho_i \in K$ for some $i$.
    As $K$ is strongly connected, but not a bottom SCC, $\rho$ may either remain within $K$ forever, or leave it.
    However, in the latter case every state reached upon leaving $K$ belongs to a trap set.
    Therefore, $\rho_{>i}$ can never leave $T'$; thus, $T'$ is a trap set.
\end{proof}

These lemmas give us a straightforward procedure (Algorithm~\ref{alg:trap}) to retrieve the smallest trap set containing a given state $q$ of an automaton $\Aut$.
At initialisation time (line~\ref{alg:trap:init1}--\ref{alg:trap:init2}), we build a directed graph $G$ out of $\Aut$, and compute its SCCs $\mathbb{S}$
as well as its condensation $\mathcal{G}$, i.e., its quotient modulo $\mathbb{S}$.
Lastly, we build a map $k : Q \rightarrow \Sccs$ that associates each state of $\Aut$ to the SCC to which it belongs.
Then, whenever we need to obtain the smallest trap set that contains state $q$, we use procedure $\textsc{MinTrapSetOf}(q)$
(lines~\ref{alg:trap:traps1}--\ref{alg:trap:traps2})
which essentially amounts to a depth-first search (DFS) on the condensation graph, rooted in $k(q)$.
Assuming that this procedure returns a list of components $K_1,\ldots,K_n$,
then $q$ belongs to the trap set $\bigcup_{i=1}^{n} K_i$. If this trap set contains the initial state of $\Aut$,
then it is trivial.
Notice that $q$ may also belong to larger trap sets, which one may find by visiting the predecessors of $k(q)$ in the condensation; however, these are not useful to our monitoring approach.
Both procedures in Algorithm~\ref{alg:trap} have linear time complexity in the worst case.

\noindent\emph{Monitoring for acceptance.}
We now describe how we can exploit the trap sets of a complete automaton $\Aut$ to give verdicts on the satisfaction of elementary acceptance conditions based on a finite prefix of a run, or at least to detect that further monitoring is hopeless.

\begin{apxtheoremrep}\label{thm:inf}
Let $T$ a trap set of \(\Aut\).
Let \(\rho\) a run of \(\Aut\) such that $\rho_i \in T$. Then,
i)~$\rho \models \Inf{S}$ for every $S \supseteq T$; and
ii)~if $T$ is minimal, then $\rho \models \Inf{S}$ for every $S$ such that $S \cap T \neq \emptyset$ and $T \setminus S$ is a transient.
\end{apxtheoremrep}
\begin{proof}
i) By Lemma~\ref{lem:trap} and since $T$ is finite,
there must exist at least one $q \in T$ that $\rho_{>i}$ visits infinitely often. But $T \subseteq S$ and therefore this state $q$ also belongs to $S$.
Thus, $\rho \not\models \Fin{S}$, which is equivalent to $\rho\models\Inf{S}$ by Def.~\ref{def:accept}.

ii) By definition of a transient, $\rho_{>i}$ leaves $T\setminus S$ infinitely often. But by Lemma~\ref{lem:trap}, it cannot leave $T$, 
and it cannot be trapped in a smaller trap set, by minimality of $T$.
Therefore, it must visit $S$ infinitely often.
\qed
\end{proof}

\begin{algorithm}[t]
\DontPrintSemicolon
\SetKwInOut{Input}{Input}\SetKwInOut{Output}{Output}
\SetKwProg{Fun}{procedure}{:}{}
\SetKwComment{Comment}{\# }{}
\Fun{$\textsc{Init}(\Aut)$}{\label{alg:trap:init1}
$G=(Q,E)\gets \text{directed graph of } \Aut$\;
$\Sccs \gets \textsc{SCC}(G)$\Comment*{strongly connected components of $G$}
\label{alg:trap:scc}
$\mathcal{G} = (\Sccs, E') \gets G/\Sccs$\Comment*{condensation graph of $G$}\label{alg:trap:cond}
$\mathit{k} \gets \{ (q, K) \in Q \times \Sccs \mid q \in K \}$
}\label{alg:trap:init2}
\Fun{$\textsc{MinTrapSetOf}(q)$}{\label{alg:trap:traps1}
    \Return $\textrm{DFS}(\mathcal{G}, k(q))$ \Comment*{Depth-first search on $\mathcal{G}$ rooted in $k(q)$}\label{alg:trap:dfs}
}\label{alg:trap:traps2}

\caption{Trap sets of an automaton $\Aut$.}\label{alg:trap}
\end{algorithm}

\begin{apxtheoremrep}\label{thm:notinf}
Let $T$ a trap set of  \(\Aut\).
Let \(\rho\) a run of $\Aut$ such that $\rho_i \in T$. Then, 
$\rho \not\models \Inf{S}$
for every $S$ such that
$S \cap T = \emptyset$.
\end{apxtheoremrep}
\begin{proof}
    Assume $\rho \models \Inf{S}$. Then there must exist $q \in S$ such that $\rho$
    visits $q$ infinitely often. Only a finite number of visits can occur in the finite prefix \(\rho_{\leq i}\). So, the remaining (infinite) visits must appear in the suffix. But $\rho_i \in T$ and, by Lemma~\ref{lem:trap}, so will every state in the suffix. But since $T$ and $S$ are disjoint, the suffix cannot visit $S$, which is a contradiction.\qed
\end{proof}

\begin{apxtheoremrep}\label{thm:ugly}
Let $T, S$ two sets of states of $\Aut$
such that $T$ is a minimal trap set,
$S \cap T \neq \emptyset$,
and such that $T\setminus S$ is not empty nor a transient.
Then every $\rho_{\leq i}$ such that
$\rho_i \in T$ is an ugly run prefix for both
$\Fin{S}$ and $\Inf{S}$.
\end{apxtheoremrep}
\begin{proof}
Assume $\rho = \rho_{\leq i}(\rho_{>i})_{\leq j}\rho_{>j}$ for some $j>i$.
Let $w=uvz$ the word for which $\rho$ is a run; that is,
let $u,v$ the finite words associated to $\rho_{\leq i}$ and $(\rho_{>i})_{\leq j}$, respectively, and $z=w_\Aut(\rho_{>j})$.

Assume $uv$ to be a good prefix for $\Fin{S}$. This means that $z$
visits $S$ only finitely often. 
Since $\rho_i \in T$, every state in $\rho_{> i}$ belongs to $T$.
By Lemma~\ref{lem:bscc1}, $T$ is strongly connected.
Thus, $\Aut$ admits a run
with prefix $\rho_{\leq i}$
such that $\rho_{>j}$ visits $S$ infinitely often, and therefore $uv$ cannot be a good prefix for $\Fin{S}$.

Additionally, it cannot be a \emph{bad} prefix for $\Fin{S}$ either,
because $T\setminus S$ is not a transient. Thus, there exists a run with prefix $\rho_{\leq i}$ such that $\rho_{>j}$ never visits $S$.
Lastly, the proof for $\Inf{S}$ follows by the fact that every good prefix of $\Fin{S}$ is a bad prefix for $\Inf{S}$, and vice versa.
\qed    
\end{proof}

Then, we can write a procedure
(Algorithm~\ref{alg:onestep}) that retrieves the trap set currently inhabited by an automaton,
and tries to determine if the execution observed so far is a good, bad, or ugly run prefix for a condition $\Inf{S}$.\footnote{%
We omit the obvious, near-identical procedure for $\Fin{S}$.}
If no verdict may be reached, the procedure returns $\bot$.
This algorithm is linear-time in the size of $\Aut$: we annotate its lines with their worst-case complexity to make this evident.
Then, a monitor repeatedly invokes this procedure after every transition of the automaton, until a conclusive verdict is reached.
By memoization, the execution time may be amortised to sublinear time.

Lastly, we generalise to compound acceptance conditions.
It is straightforward to use the following theorems as a basis to build composite monitors.

\begin{apxtheoremrep}
Let $\phi = \varphi_1 \land \varphi_2$ an acceptance condition.
If $\rho_{\leq i}$ is a bad or ugly run prefix for at least one of $\varphi_{1,2}$, then $\rho_{\leq i}$ is a bad or ugly run prefix for $\phi$.
If $\rho_{\leq i}$ is a good prefix for both $\varphi_1$ and $\varphi_2$, then it is a good run prefix for $\phi$.
\end{apxtheoremrep}
\begin{proof}
If $\rho_{\leq i}$ is a bad prefix for $\varphi_1$, we can conclude that $\rho \not\models \varphi_1$: thus, no suffix of $\rho_{\leq i}$ can produce a word that models $\varphi_1$.
But $\rho\models\varphi_1$ is a requirement to $\rho\models\phi$, therefore $\rho_{\leq i}$ is a bad prefix for $\phi$.

If $\rho_{\leq i}$ is an ugly run prefix for $\varphi_1$, then no further finite monitoring can determine whether $\rho$ models $\varphi_1$. Again, reaching a verdict on this is necessary to determine whether $\rho$ models $\phi$, therefore $\rho$ is an ugly run prefix for~$\phi$.

If $\rho_{\leq i}$ is a good run prefix for both sub-conditions, then $\rho$ models both conditions. This implies that it models their conjunction as well.
\end{proof}

\begin{apxtheoremrep}
Let $\phi = \varphi_1 \lor \varphi_2$ an acceptance condition.
If $\rho_{\leq i}$ is a good run prefix for $\varphi_{1}$ or $\varphi_2$, then $\rho_{\leq i}$ is a good run prefix for $\phi$.
If $\rho_{\leq i}$ is a bad run prefix for both $\varphi_{1,2}$, then it is a bad run prefix for $\phi$.
If $\rho_{\leq i}$ is an ugly run prefix for one
of $\varphi_{1,2}$, and bad or ugly for the other, then it is an ugly run prefix for $\phi$.
\end{apxtheoremrep}
\begin{proof}
    As soon as we can determine that $\rho$ models one of the two sub-conditions, we can conclude that it models their disjunction as well. Thus, we only need $\rho_{\leq i}$ to be a good run prefix for either $\varphi_1$ or $\varphi_2$.
    Conversely, when we encounter a bad run prefix for, say, $\varphi_1$, then we have to keep monitoring for $\varphi_2$ (and vice versa) in order to reach a verdict on the disjunction.
    If we have a bad run prefix $\rho_{\leq i}$ for both sub-conditions, then we can conclude that $\rho$ does not model either of them, and thus does not model $\phi$ either.
    Lastly, if $\rho_{\leq i}$ is a bad prefix for one disjunct and an ugly prefix for the other, then no further monitoring can lead us to conclude whether $\rho$ models the latter. Thus, we cannot conclude anything about the disjunction, i.e., the prefix is an ugly run prefix for $\phi$.
\end{proof}

\begin{algorithm}[t]
\DontPrintSemicolon
\SetKwInOut{Input}{Input}\SetKwInOut{Output}{Output}
\SetKwComment{Comment}{\# }{}
\Input{A deterministic complete automaton $\Aut$ and its current state $q$}
\Output{\textsf{good}, \textsf{bad}, \textsf{ugly}, or $\bot$.}
\SetKwComment{Comment}{\# }{}
    $\textit{Comps} \gets $  $\textsc{MinTrapSetOf}(q)$\Comment*{$O(|Q|)$ (Algorithm~\ref{alg:trap})}
    $q_0 \gets$ initial state of $\Aut$\;
    $T \gets \bigcup \textit{Comps}$\Comment*{$O(|Q|)$ (components are disjoint)}
    \lIf(\Comment*[f]{$O(|S|)$ (Theorem~\ref{thm:inf})}){$T \subseteq S$}{\Return \textsf{good}}
    \lIf(\Comment*[f]{$O(|S|)$ (Theorem~\ref{thm:notinf})}){$T \cap S =\emptyset$}{\Return \textsf{bad}}
    \If(\Comment*[f]{if $T$ is minimal (Theorems~\ref{thm:inf} and~\ref{thm:ugly})}){$|\textit{Comps}| = 1$}{
        \leIf(\Comment*[f]{$O(|S|+|E|)$}){$T \setminus S$ is transient}{\Return \textsf{good}}{\Return \textsf{ugly}}}
    \Return $\bot$

    \caption{One-step monitor for acceptance condition $\Inf{S}$.}\label{alg:onestep}
\end{algorithm}

\section{A Tool for Executing HOA $\omega$-Automata}\label{sec:tool}

In this section we introduce and evaluate our tool, \HOAexec.
The tool is implemented in Python, it is open-source
and available online.\footnote{\url{https://github.com/lou1306/hoax}.
A persistent snapshot is also available~\cite{luca_di_stefano_2025_15909340}.}

\noindent\emph{Usage and architecture.}
The tool accepts as input one or more HOA automata and a configuration file.
The HOA files are used to construct one or more \emph{runners} that will keep track of each automaton's state.
The configuration file, in turn, describes how inputs to the automata should be retrieved or generated. 
Each way of providing inputs to automata is reified by means of \emph{driver} objects.
At the moment, \HOAexec provides drivers that implement an interactive user prompt, a text file reader, or pseudo-random Boolean generators with user-defined biases. Different atomic propositions may be mapped to different drivers, and the user may declare a default driver for propositions that remain unmapped.

In the main execution loop, \HOAexec collects a valuation of $\AAPP$ by interrogating the drivers, and feeds it to the runners which choose a suitable next state for their respective automata. Runners may also be equipped with reactive \emph{hooks}. A hook is a combination of a \emph{trigger} object that monitors whether the automaton has met a certain condition, and an \emph{action} that is invoked whenever this happens.
For instance, the user may use hooks to specify how the runner should resolve nondeterminism, or how it should behave on deadlock. Possible actions for these triggers include choosing a random option, prompting the user, resetting the automaton, or forcing it into a specific state.
In fact, we have implemented the monitoring approach described in this work as a family of triggers that activate a reset action whenever Algorithm~\ref{alg:onestep} gives a conclusive verdict.\footnote{The assumption here is that, whenever a violation is detected, some external mechanism brings the environment back to a safe state.}

\noindent\emph{Experimental evaluation.}%
We consider a lock acquisition scenario from~\cite{DBLP:conf/rv/DamsHK22}, where $N$ threads compete to acquire $N$ locks.
We observe finite traces of events in the form $a_{i\ell}$ and $r_{i\ell}$, indicating that thread $i$ has acquired or released lock $\ell$.
An additional event $\textit{end}$ denotes the end of a trace.
We want to monitor that no lock acquired by $i$ is then acquired by $j\neq i$ before being released~\eqref{eq:nodoubleacq}; and that, by the end of the trace, all acquired locks have been released~\eqref{eq:releasebyend}. Acquisition events that occur at the end of the trace are ignored.

\newenvironment{talign}
 {\let\displaystyle\textstyle\align}
 {\endalign}

{
\begin{talign}
\textsc{NoDoubleAcq}_{i\ell} &\triangleq G (\neg\textit{end} \land a_{i\ell} \implies r_{i\ell} \mathrel{R} \bigwedge_{j\neq i} \neg a_{j\ell})\label{eq:nodoubleacq}\\
\textsc{ReleaseByEnd}_{i\ell} &\triangleq G (\neg\textit{end} \land a_{i\ell} \implies \neg\textit{end} \mathrel{W} r_{j\ell})\label{eq:releasebyend}
\end{talign}
}

\pgfplotsset{
    discard if not/.style 2 args={
        /pgfplots/boxplot/data filter/.code={
            \edef\tempa{\thisrow{#1}}
            \edef\tempb{#2}
            \ifx\tempa\tempb
            \else
                \def\pgfmathresult{inf}
            \fi
        }
    }
}
\pgfplotsset{
    only if/.style args={entry of #1 is #2}{
        /pgfplots/boxplot/data filter/.code={
            \edef\tempa{\thisrow{#1}}
            \edef\tempb{#2}
            \ifx\tempa\tempb
            \else
                \def\pgfmathresult{}
            \fi
        }
    }
}
\pgfplotsset{
    only if snd/.style args={entry of #1 is #2}{
        /pgfplots/boxplot/data filter/.code={
            \edef\tempa{\thisrow{#1}}
            \edef\tempb{#2}
            \ifx\tempa\tempb
            \else
                \def\pgfmathresult{}
            \fi
        }
    }
} 

\newcommand{\EX}[1]{\!\!\times\!\!10^{#1}}

\begin{figure}[t]
    \centering
    \scriptsize
\begin{tikzpicture}
\begin{axis}[
    boxplot,
    width=\textwidth,
    height=0.3\textwidth,
    table/y=time_seconds,
    boxplot/draw direction=x,
    ylabel = {Trace length},
    xmin=0, xmax=4.1,
    ymin=0.5, ymax=6.5,
    ytick={1.5,3.5,5.5},
        ytick style={draw=none},
    yticklabels={{$5\EX{4}$},{$1\EX{5}$},{$2\EX{5}$}},
    legend to name={legend},
    name=border
]
    
    \addplot[draw=black, mark=o] table[discard if not={tool}{hoax},col sep=comma]{N2L50.csv};
    \addplot[pattern = north east lines, pattern color=red, mark=o] table[discard if not={tool}{pycontract},col sep=comma]{N2L50.csv};
    \addplot[draw=black, mark=o] table[discard if not={tool}{hoax},col sep=comma]{N2L100.csv};
    \addplot[pattern = north east lines, pattern color=red, mark=o] table[color=red, discard if not={tool}{pycontract},col sep=comma]{N2L100.csv};
    \addplot[draw=black, mark=o] table[discard if not={tool}{hoax},col sep=comma]{N2L200.csv};
    \addplot[pattern = north east lines, pattern color=red, mark=o] table[discard if not={tool}{pycontract},col sep=comma]{N2L200.csv};
    \addplot [xcomb, draw=gray] coordinates {(18,2.5)};
    \addplot [xcomb, draw=gray] coordinates {(18,4.5)};
\end{axis}
\node[above left,fill=white] at (1,1.6) {$N=2$};
\end{tikzpicture}

\begin{tikzpicture}
\begin{axis}[
    boxplot,
    width=\textwidth,
    height=0.3\textwidth,
    table/y=time_seconds,
    boxplot/draw direction=x,
    ylabel = {Trace length},
    xmin=0, xmax=7.1,
    ymin=0.5, ymax=6.5,
    ytick={1.5,3.5,5.5},
    ytick style={draw=none},
    yticklabels={{$5\EX{4}$},{$1\EX{5}$},{$2\EX{5}$}},
    legend to name={legend},
    name=border
]
    \addplot[draw=black, mark=o] table[discard if not={tool}{hoax},col sep=comma]{N4L50.csv};
    \addplot[pattern = north east lines, pattern color=red, mark=o] table[discard if not={tool}{pycontract},col sep=comma]{N4L50.csv};
    \addplot[draw=black, mark=o] table[discard if not={tool}{hoax},col sep=comma]{N4L100.csv};
    \addplot[pattern = north east lines, pattern color=red, mark=o] table[color=red, discard if not={tool}{pycontract},col sep=comma]{N4L100.csv};
    \addplot[draw=black, mark=o] table[discard if not={tool}{hoax},col sep=comma]{N4L200.csv};
    \addplot[pattern = north east lines, pattern color=red, mark=o] table[discard if not={tool}{pycontract},col sep=comma]{N4L200.csv};
    \addplot [xcomb, draw=gray] coordinates {(18,2.5)};
    \addplot [xcomb, draw=gray] coordinates {(18,4.5)};
\end{axis}
\node[above left,fill=white] at (1,1.6) {$N=4$};
\end{tikzpicture}

\begin{tikzpicture}
\begin{axis}[
    width=\textwidth,
    height=0.3\textwidth,
    ylabel = {Trace length},
    xlabel = {Execution time (s)},
    xmin=0, xmax=16.5,
    ymin=0.5, ymax=6.5,
    ytick={1.5,3.5,5.5},
    ytick style={draw=none},
    yticklabels={{$5\EX{4}$},{$1\EX{5}$},{$2\EX{5}$}},
    legend columns = 2,
    legend to name={legend},
    name=border
]
    \addplot[boxplot, table/y=time_seconds, boxplot/draw direction=x,draw=black, mark=o, area legend] table[discard if not={tool}{hoax},col sep=comma]{N8L50.csv};
    \addlegendentry[mark=o]{\HOAexec};
    \addplot[boxplot, table/y=time_seconds, boxplot/draw direction=x,pattern = north east lines, pattern color=red, mark=o, area legend] table[discard if not={tool}{pycontract},col sep=comma]{N8L50.csv};
    \addlegendentry[mark={o, fill=red}]{PyContract};
    \addplot[boxplot, table/y=time_seconds, boxplot/draw direction=x,draw=black, mark=o] table[discard if not={tool}{hoax},col sep=comma]{N8L100.csv};
    \addplot[boxplot, table/y=time_seconds, boxplot/draw direction=x,pattern = north east lines, pattern color=red, mark=o] table[color=red, discard if not={tool}{pycontract},col sep=comma]{N8L100.csv};
    \addplot[boxplot, table/y=time_seconds, boxplot/draw direction=x,draw=black, mark=o] table[discard if not={tool}{hoax},col sep=comma]{N8L200.csv};
    \addplot[boxplot, table/y=time_seconds, boxplot/draw direction=x,pattern = north east lines, pattern color=red] table[discard if not={tool}{pycontract},col sep=comma]{N8L200.csv};
    \addplot [xcomb, draw=gray] coordinates {(18,2.5)};
    \addplot [xcomb, draw=gray] coordinates {(18,4.5)};
\end{axis}
\node[below left, yshift=-8pt,xshift=1pt] at (border.south east) {\ref{legend}};
\node[above left,fill=white] at (1,1.6) {$N=8$};
\end{tikzpicture}
\caption{Running times of \HOAexec and PyContract on the lock acquisition systems.}\label{fig:experiments}
\vspace*{-2em}
\end{figure}
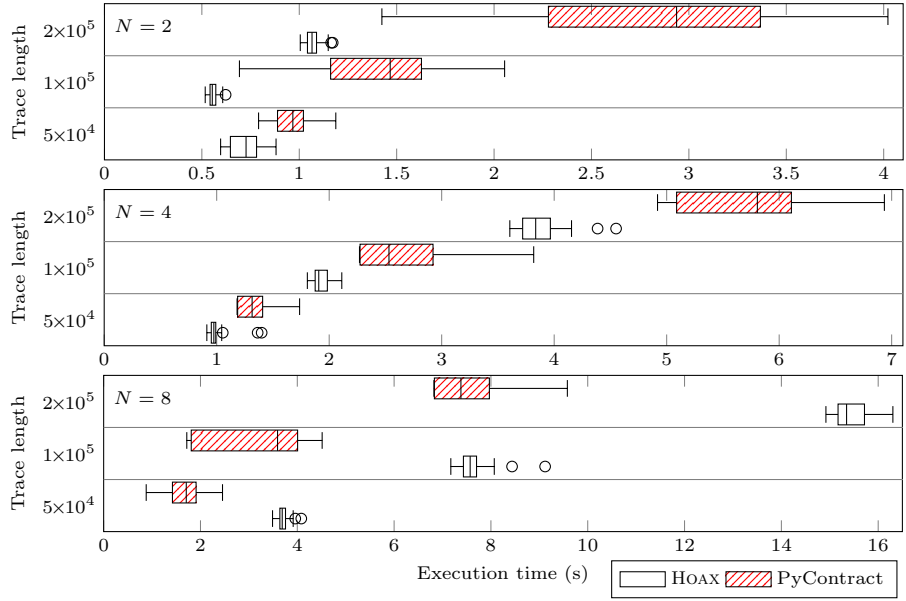
 
We compare how PyContract and \HOAexec deal with this scenario.
For PyContract, we reuse monitor \texttt{M1} from~\cite{DBLP:conf/rv/DamsHK22}. For \HOAexec, we use Spot to generate automata from the negation of the above properties for each pair $(i, \ell)$.
We consider three systems, respectively with $N=2,4,8$, and generate traces of length $L=5\times10^4, 1\times10^5,2\times10^5$ for each system.
Notice that, to encode events $a_{i\ell}$ and $r_{i\ell}$ in \HOAexec traces, we use a single proposition \texttt{a} to denote whether the event is an acquisition (when asserted) or a release (when negated), and an additional $2\log_2 N$ variables that binary-encode the values of $i$ and $\ell$.
We ran each tool 50 times on each pair $(N, L)$. Experiments were performed on an Apple MacBook Pro laptop with an M3 Pro CPU and 18 GiB of RAM, running macOS 14.7.3.\footnote{We provide a repository at \url{https://github.com/lou1306/hoax-experiments/} containing the 
code used for these experiments, to facilitate replication.
A persistent snapshot is also available~\cite{di_stefano_2025_15908703}.
We used this repository to replicate our experiments on an Intel i7-5820K workstation with 32~GiB of RAM, running Ubuntu 22.04.}

Fig.~\ref{fig:experiments} reports box plots for all running times on the three systems. We only track the time that the tools spend analysing the trace and reporting violations, and ignore all setup operations (instantiating the \texttt{M1} class for PyContract, and parsing the automata files for \HOAexec).
On smaller systems ($N=2,4$), \HOAexec shows faster execution times, with a smaller variance and a lower sensitivity to the length of the trace.
However, PyContract is faster in the larger setting, $N=8$.
Note that properties (\ref{eq:nodoubleacq}--\ref{eq:releasebyend}) correspond to $2N^2$ (deterministic and complete) automata on a system of size $N$. The automata all have the same size, i.e., 3 states and 6 transitions, but their set of APs grows as $N$ increases ($2+2\log_2 N$).
Therefore, when $N=8$, \HOAexec deals with 128 automata over 8 atomic propositions, while PyContract's use of one symbolic monitor
starts to pay off.

\section{Conclusions}\label{sec:conclusion}

We have introduced \HOAexec, an open-source execution tool for HOA automata that supports monitoring of arbitrary acceptance conditions.
We have remarked that runs of a nonmonitorable automaton may not be recognized up to any finite prefix, and shown that our tool may at least be able to recognize ugly prefixes and give up hopes of recognizing the current run.
The tool is highly configurable, and its modular architecture should facilitate future extensions.
Although \HOAexec was not designed to compete with
symbolic trace analysis, our experimental comparison with PyContract looks promising.
At the moment, automata are executed sequentially, and parallelisation might substantially improve performance.
The main advantages of our tool, in any case, are its configurability and interoperability with the rich collection of
existing utilities for HOA generation and analysis,
allowing to directly create monitors from LTL specifications
rather than having to program them manually.

Support for nondeterminism would be an attractive direction for future work,
allowing for smaller automata than their deterministic counterparts (when they exist).
After a run prefix, these automata may be in one of multiple states, possibly across disjoint trap sets, and
we would need to consider all these sets before reaching a conclusion on the quality of the prefix.
Likely, this would entail tracking them symbolically, to palliate state space explosion.
Automata formats that can embed more general logics, such as LTL modulo theories~\cite{DBLP:conf/cav/RodriguezS23}, may partially overcome the scalability limitations we have observed in our experimental evaluation by enabling monitors with richer inputs, such as integers.
To fully match the power of symbolic approaches such as PyContract, one would need richer, data-aware logics, such as Constraint LTL~\cite{DBLP:journals/iandc/DemriD07} or TSL~\cite{DBLP:conf/cav/Finkbeiner0PS19,DBLP:conf/fossacs/FinkbeinerHP22}.

\bibliographystyle{splncs04}
\bibliography{biblio}

\begin{thebibliography}{10}
\providecommand{\url}[1]{\texttt{#1}}
\providecommand{\urlprefix}{URL }
\providecommand{\doi}[1]{https://doi.org/#1}

\bibitem{DBLP:conf/cav/BabiakBDKKM0S15}
Babiak, T., Blahoudek, F., Duret-Lutz, A., Klein, J., Kretínský, J., Müller, D., Parker, D., Strejcek, J.: The {Hanoi} {Omega}-{Automata} {Format}. In: 27th {International} {Conference} on {Computer} {Aided} {Verification} ({CAV}). {LNCS}, vol.~9206, pp. 479--486. Springer (2015). \doi{10.1007/978-3-319-21690-4\_31}

\bibitem{DBLP:journals/tosem/BauerLS11}
Bauer, A., Leucker, M., Schallhart, C.: Runtime {Verification} for {LTL} and {TLTL}. ACM Transactions on Software Engineering and Methodology  \textbf{20}(4),  14:1--14:64 (2011). \doi{10.1145/2000799.2000800}

\bibitem{biere2007}
Biere, A.: The {{AIGER And-Inverter Graph}} ({{AIG}}) {{Format Version}} 20071012. Technical {{Report}}, Johannes Kepler University (2007), \url{https://doi.org/10.35011/fmvtr.2007-1}

\bibitem{DBLP:conf/tacas/BunteGKLNVWWW19}
Bunte, O., Groote, J.F., Keiren, J.J.A., Laveaux, M., Neele, T., de~Vink, E.P., Wesselink, W., Wijs, A., Willemse, T.A.C.: The {mCRL2} toolset for analysing concurrent systems - improvements in expressivity and usability. In: 25th {International} {Conference} on {Tools} and algorithms for the construction and analysis of systems ({TACAS}). {LNCS}, vol. 11428, pp. 21--39. Springer (2019). \doi{10.1007/978-3-030-17465-1\_2}

\bibitem{DBLP:conf/rv/DamsHK22}
Dams, D., Havelund, K., Kauffman, S.: A {{Python Library}} for {{Trace Analysis}}. In: 22nd {{International Conference}} on {{Runtime Verification}} ({{RV}}). {{LNCS}}, vol. 13498, pp. 264--273. Springer (2022). \doi{10.1007/978-3-031-17196-3\_15}

\bibitem{DBLP:journals/iandc/DemriD07}
Demri, S., D'Souza, D.: An automata-theoretic approach to constraint {{LTL}}. Inf. Comput.  \textbf{205}(3),  380--415 (2007). \doi{10.1016/j.ic.2006.09.006}

\bibitem{luca_di_stefano_2025_15909340}
Di~Stefano, L.: Lou1306/hoax: V0.1.1. Zenodo (2025). \doi{pv9x}

\bibitem{di_stefano_2025_15908703}
Di~Stefano, L.: Software artifact for ``{{Execution}} and monitoring of {{HOA}} automata with {{HOAX}}''. Zenodo (2025). \doi{10.5281/zenodo.15908703}

\bibitem{DBLP:conf/atva/Duret-LutzLFMRX16}
Duret-Lutz, A., Lewkowicz, A., Fauchille, A., Michaud, T., Renault, E., Xu, L.: Spot 2.0 - {A} framework for {LTL} and $\omega$-{Automata} manipulation. In: 14th international symposium on {Automated} technology for verification and analysis ({ATVA}). {LNCS}, vol.~9938, pp. 122--129 (2016). \doi{10.1007/978-3-319-46520-3\_8}

\bibitem{DBLP:conf/fossacs/FinkbeinerHP22}
Finkbeiner, B., Heim, P., Passing, N.: Temporal {{Stream Logic}} modulo {{Theories}}. In: 25th {{International Conference}} on {{Foundations}} of {{Software Science}} and {{Computation Structures}} ({{FOSSACS}}). {{LNCS}}, vol. 13242, pp. 325--346. Springer (2022). \doi{10.1007/978-3-030-99253-8\_17}

\bibitem{DBLP:conf/cav/Finkbeiner0PS19}
Finkbeiner, B., Klein, F., Piskac, R., Santolucito, M.: Temporal {{Stream Logic}}: {{Synthesis Beyond}} the {{Bools}}. In: 31st {{International Conference}} on {{Computer Aided Verification}} ({{CAV}}). {{LNCS}}, vol. 11561, pp. 609--629. Springer (2019). \doi{10.1007/978-3-030-25540-4\_35}

\bibitem{DBLP:journals/sttt/GaravelLMS13}
Garavel, H., Lang, F., Mateescu, R., Serwe, W.: {{CADP}} 2011: {{A}} toolbox for the construction and analysis of distributed processes. Software Tools for Technology Transfer  \textbf{15}(2),  89--107 (2013). \doi{10.1007/s10009-012-0244-z}

\bibitem{DBLP:conf/birthday/GaravelLS17}
Garavel, H., Lang, F., Serwe, W.: From {LOTOS} to {LNT}. In: {ModelEd}, {TestEd}, {TrustEd} - {Essays} {Dedicated} to {Ed} {Brinksma} on the {Occasion} of {His} 60th {Birthday}. {LNCS}, vol. 10500, pp. 3--26. Springer (2017). \doi{10.1007/978-3-319-68270-9\_1}

\bibitem{DBLP:journals/nc/KlarnerBS15}
Klarner, H., Bockmayr, A., Siebert, H.: Computing maximal and minimal trap spaces of {{Boolean}} networks. Natural Computing  \textbf{14}(4),  535--544 (2015). \doi{10.1007/s11047-015-9520-7}

\bibitem{DBLP:conf/atva/KretinskyMS18}
Kretínský, J., Meggendorfer, T., Sickert, S.: Owl: {A} {Library} for $\omega$-{Words}, {Automata}, and {LTL}. In: 16th {International} {Symposium} on {Automated} {Technology} for {Verification} and {Analysis} ({ATVA}). {LNCS}, vol. 11138, pp. 543--550. Springer (2018). \doi{10.1007/978-3-030-01090-4\_34}

\bibitem{DBLP:journals/fmsd/KupfermanV01}
Kupferman, O., Vardi, M.Y.: Model {{Checking}} of {{Safety Properties}}. Formal Methods in System Design  \textbf{19}(3),  291--314 (2001). \doi{10.1023/A:1011254632723}

\bibitem{DBLP:conf/cav/KwiatkowskaNP11}
Kwiatkowska, M.Z., Norman, G., Parker, D.: {PRISM} 4.0: {Verification} of {Probabilistic} {Real}-{Time} {Systems}. In: 23rd {International} {Conference} on {Computer} {Aided} {Verification} ({CAV}). {LNCS}, vol.~6806, pp. 585--591. Springer (2011). \doi{10.1007/978-3-642-22110-1\_47}

\bibitem{DBLP:conf/focs/Pnueli77}
Pnueli, A.: The {{Temporal Logic}} of {{Programs}}. In: 18th {{Annual Symposium}} on {{Foundations}} of {{Computer Science}} ({{FOCS}}). pp. 46--57. IEEE (1977). \doi{10.1109/SFCS.1977.32}

\bibitem{DBLP:conf/cav/RodriguezS23}
Rodríguez, A., Sánchez, C.: Boolean {Abstractions} for {Realizability} {Modulo} {Theories}. In: 35th {International} {Conference} on {Computer} {Aided} {Verification} ({CAV}). {LNCS}, vol. 13966, pp. 305--328. Springer (2023). \doi{10.1007/978-3-031-37709-9\_15}

\bibitem{DBLP:conf/charme/YuML99}
Yu, Y., Manolios, P., Lamport, L.: Model checking {TLA}+ specifications. In: 10th advanced research working conference on correct hardware design and verification methods ({CHARME}). {LNCS}, vol.~1703, pp. 54--66. Springer (1999). \doi{10.1007/3-540-48153-2\_6}

\end{thebibliography}

\end{document}